\documentclass{IEEEtran4PSCC}
\ifCLASSINFOpdf
   \usepackage[pdftex]{graphicx}
\else
   \usepackage[dvips]{graphicx}
\fi
\makeatletter
\let\old@ps@headings\ps@headings
\let\old@ps@IEEEtitlepagestyle\ps@IEEEtitlepagestyle
\def\psccfooter#1{%
    \def\ps@headings{%
        \old@ps@headings%
        \def\@oddfoot{\strut\hfill#1\hfill\strut}%
        \def\@evenfoot{\strut\hfill#1\hfill\strut}%
    }%
    \def\ps@IEEEtitlepagestyle{%
        \old@ps@IEEEtitlepagestyle%
        \def\@oddfoot{\strut\hfill#1\hfill\strut}%
        \def\@evenfoot{\strut\hfill#1\hfill\strut}%
    }%
    \ps@headings%
}
\makeatother

\psccfooter{%
        \parbox{\textwidth}{\hrulefill \\ \small{24th Power Systems Computation Conference} \hfill \begin{minipage}{0.2\textwidth}\centering \vspace*{4pt} \includegraphics[scale=0.06]{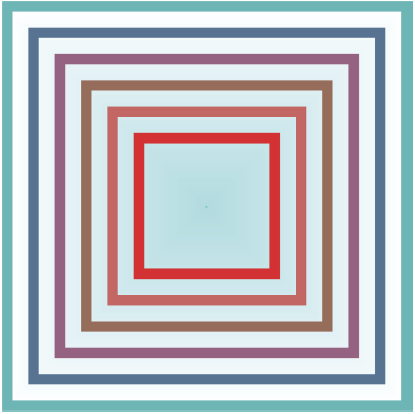}\\\small{PSCC 2026} \end{minipage} \hfill \small{Limassol, Cyprus --- June 8 -- June 12, 2026}}%
}
\usepackage[cmex10]{amsmath}
\usepackage{amssymb}
\usepackage{bm}
\usepackage{stmaryrd}
\usepackage{xcolor}
\usepackage{acro}
\usepackage{hyperref}
\usepackage{tikz, pgfplots}
\pgfplotsset{compat=1.18}
\usetikzlibrary{shapes.geometric, shapes.arrows, arrows.meta, calc, positioning, fit, pgfplots.groupplots, backgrounds}
\usepgfplotslibrary{fillbetween,statistics}
\usepackage{algorithm}
\usepackage{algorithmicx}
\usepackage{algpseudocode}
\usepackage[
    backend=biber,
    style=ieee,
    sorting=none
  ]{biblatex}
\usepackage{amsthm}
\newtheorem{theorem}{Theorem}
\newtheorem{definition}{Definition}

\usepackage[final]{changes}

\usepackage{ifthen}
\usepackage{xparse}

\newcommand{\val}[1]{
    v_{#1}
}
\newcommand{\lval}[1]{
    \hat{v}_{#1}
}
\newcommand{\params}[2]{
    \theta_{#1}^{#2}
}
\newcommand{\realN}[1]{
    \mathbb{R}^{#1}
}
\newcommand{\packageset}[1]{
    \mathcal{X}_{#1}
}

\newcommand{\npros}{
    n
}
\newcommand{\nprod}{
    m
}
\newcommand{\package}[2]{
    \mathbf{x}_{#1}^{#2}
}
\newcommand{\product}[2]{
    x_{#1}^{#2}
}

\newcommand{\dpackage}[2]{
    \tilde{\mathbf{x}}_{#1}^{#2}
}
\newcommand{\lpackage}[2]{
    \hat{\mathbf{x}}_{#1}^{#2}
}

\newcommand{\linprice}[2]{
    \boldsymbol{\lambda}_{#1}^{#2}
}
\newcommand{\price}[2]{
    \lambda_{#1}^{#2}
}

\newcommand{\linpricelow}[2]{
    \underline{\boldsymbol{\lambda}}_{#1}^{#2}
}
\newcommand{\linpricehigh}[2]{
    \bar{\boldsymbol{\lambda}}_{#1}^{#2}
}
\newcommand {\horizon}{
    T
}
\newcommand{\lagrangedual}{
    g
}
\newcommand{\llagrangedual}{
    \hat{g}
}
\newcommand{\lrcce}[1]{
    \eta_{#1}
}
\newcommand{\lrmlcce}[1]{
    \hat{\eta}_{#1}
}
\newcommand{\dataset}[2]{
    \mathcal{D}_{#1}^{#2}
}
\newcommand{\intset}[2]{
    \{{#1}, \dots, {#2}\}
}
\newcommand{\maxpackage}[1]{
    \bar{x}_{#1}
}

\newcommand{\soutput}[2]{
    y_{#1,#2}
}
\newcommand{\voutput}[1]{
    \mathbf{y}_{#1}
}

\NewDocumentCommand{\soe}{g}{
    \IfNoValueTF{#1}
        {\mathbf{s}}
        {s_{#1}}
}

\NewDocumentCommand{\soer}{g}{
    \IfNoValueTF{#1}
        {\hat{\mathbf{s}}}
        {\hat{s}_{#1}}
}
\newcommand{\soepenalty}[1]{
    \boldsymbol{\alpha}_{#1}
}
\newcommand{\soecap}[0]{
    \bar{s}
}

\newcommand*{\horizonset}{\mathcal{H}}
\newcommand*{\indexset}[1]{\mathcal{I}_{#1}}

\newcommand*{\fixedprodnot}[0]{f}
\newcommand*{\storagenot}[0]{s}
\newcommand*{\defloadnot}[0]{d}

\DeclareAcronym{SOE}{
    short = SoE,
    long = state of energy,
    tag = abbrev,
    long-plural-form = states of energy
}
\DeclareAcronym{EV}{
    short = EV,
    long = electric vehicle,
    tag = abbrev
}
\DeclareAcronym{DER}{
    short = DER,
    long = distributed energy resource,
    tag = abbrev
}
\DeclareAcronym{ML}{
    short = ML,
    long = machine learning,
    tag = abbrev
}
\DeclareAcronym{MVNN}{
    short = MVNN,
    long = monotone valued neural network,
    tag = abbrev
}
\DeclareAcronym{IO}{
    short = IO,
    long = inverse optimization,
    tag = abbrev
}
\DeclareAcronym{CHP}{
    short = CHP,
    long = convex hull pricing,
    tag = abbrev
}
\DeclareAcronym{LMP}{
    short = LMP,
    long = locational marginal pricing,
    tag = abbrev
}
\DeclareAcronym{LEM}{
    short = LEM,
    long = local energy market,
    tag = abbrev
}
\DeclareAcronym{IR}{
    short = IR,
    long = individual rationality,
    tag = abbrev
}
\DeclareAcronym{IC}{
    short = IC,
    long = incentive compatibility,
    tag = abbrev
}
\DeclareAcronym{BB}{
    short = BB,
    long = budget balanced,
    tag = abbrev
}
\DeclareAcronym{VCG}{
    short = VCG,
    long = Vickrey-Clarke-Groves,
    tag = abbrev
}
\DeclareAcronym{DSO}{
    short = DSO,
    long = distribution system operator,
    tag = abbrev
}

\DeclareAcronym{MILP}{
    short = MILP,
    long = mixed-integer linear program,
    tag = abbrev
}
\DeclareAcronym{CCE}{
    short = CCE,
    long = combinatorial clock exchange,
    tag = abbrev
}
\DeclareAcronym{MLCCE}{
    short = MLCCE,
    long = ML-aided combinatorial clock exchange,
    tag = abbrev
}

\DeclareAcronym{dispatch}{
    short = \(\dispatch{i}{h}\),
    long = Dispatch of device \(i\) in hour \(h\),
    tag = var
}
\DeclareAcronym{stviol}{
    short = $\stviolup{i}{h}\text{, }\stviollow{i}{h}$,
    long = SoE upper and lower limit violation of device \(i\) in hour \(h\),
    tag = var
}
\DeclareAcronym{stcycvar}{
    short = \(\stcycvar{i}{h}\),
    long = Amount of storage cycled for device \(i\) in hour \(h\),
    tag = var
}
\DeclareAcronym{dlvars}{
    short = \(\dlhourvar{i}{h}\text{, }\dlvar{i}\),
    long = Auxilliary variables for device \(i\) and hour \(h\),
    tag = var
}

\DeclareAcronym{prodlowup}{
    short = \(\produp{i}{h}\text{, }\prodlow{i}{h}\),
    long = Production limits for fixed production device \(i\) and hour \(h\),
    tag = par
}
\DeclareAcronym{soeuplow}{
    short = \(\soeup{i}{h}\text{, }\soelow{i}{h}\),
    long = SoE limits for storage device \(i\) and hour \(h\),
    tag = par
}
\DeclareAcronym{penstvioluplow}{
    short = \(\penstviolup{i}{h}\text{, }\penstviollow{i}{h}\),
    long = Penaly parameters for SoE limits violation for storage device \(i\) and hour \(h\),
    tag = par
}
\DeclareAcronym{linutcoef}{
    short = \(\linutcoef{i}{h}\),
    long = Utility parameters for fixed production device \(i\) and hour \(h\),
    tag = par
}
\DeclareAcronym{effcoef}{
    short = \(\effcoef{i}\),
    long = Coefficient of charging/discharging efficiency for storage device \(i\),
    tag = par
}
\DeclareAcronym{disscoef}{
    short = \(\disscoef{i}\),
    long = Rate of energy dissipation for storage device \(i\),
    tag = par
}
\DeclareAcronym{degrcoef}{
    short = \(\degrcoef{i}\),
    long = Degradation parameter for storage device \(i\) and hour \(h\),
    tag = par
}
\DeclareAcronym{dltimeuplow}{
    short = \(\dltimeup{i}\text{, }\dltimelow{i}\),
    long = Completion time limits for deferrable load \(i\),
    tag = par
}
\DeclareAcronym{dltimereq}{
    short = \(\dltimereq{i}\),
    long = Time required for load completion for deferrable load \(i\),
    tag = par
}
\DeclareAcronym{dlload}{
    short = \(\dlload{i}\),
    long = Power required for deferrable load \(i\),
    tag = par
}
\DeclareAcronym{dlpenalty}{
    short = \(\dlpenalty{i}\),
    long = Penalty for unsuccessful completion of deferrable load \(i\),
    tag = par
}
\DeclareAcronym{horizonset}{
    short = \(\horizonset\),
    long = Set of time intervals in the planning horizon,
    tag = notation
}
\DeclareAcronym{indexset}{
    short = \(\indexset{q}\),
    long = Set of indices for device type \(q\),
    tag = notation
}
\DeclareAcronym{devicenot}{
    short = \(\fixedprodnot\text{, }\storagenot\text{, }\defloadnot\),
    long = Device type notations,
    tag = notation
}
\begin{document}
%


\title{Simplifying Preference Elicitation in Local Energy Markets: Combinatorial Clock Exchange}

\author{
\IEEEauthorblockN{Shobhit Singhal and Lesia Mitridati}
\IEEEauthorblockA{Department of Wind and Energy Systems \\
Technical University of Denmark, Kgs. Lyngby, Denmark\\
\{shosi, lemitri\}@dtu.dk}
}


\maketitle

\begin{abstract}
As \acp*{DER} proliferate, future power system will need new market platforms enabling prosumers to trade various electricity and grid-support products. 
However, prosumers often exhibit complex, product interdependent preferences and face limited cognitive capacity, hindering participation in prevailing markets with complex structures and bid formats. We address this challenge by introducing a multi-product market that allows prosumers to express complex preferences through an intuitive format, by fusing combinatorial clock exchange and machine learning (ML) techniques. The iterative mechanism only requires prosumers to report their preferred package of products at posted prices, eliminating the need for forecasting product prices or adhering to complex bid formats, while the ML-aided price discovery speeds up convergence. The linear pricing rule further enhances transparency and interpretability. Finally, numerical simulations demonstrate convergence to clearing prices in approximately $15$ clock iterations.
\end{abstract}

\begin{IEEEkeywords}
combinatorial exchange, local energy markets, machine learning, prosumers, Walrasian equilibrium.\thanksto{\noindent Submitted to the 24th Power Systems Computation Conference (PSCC 2026).}
\end{IEEEkeywords}

\section{Introduction}\label{sec:intro}

The power system is witnessing a surge of \acp{DER}, such as rooftop solar PV, \acp{EV}, and data centers. These resources can draw or inject huge amounts of power, posing threat to grid stability. However, many of these resources are flexible and are potentially valuable for managing the stability of future renewable-dominated power systems. Furthermore, advancements in information and communications technology (ICT) and energy management systems enable \ac{DER} owners, so-called \emph{prosumers}, to trade electricity and grid-support services, such as frequency regulation, reactive power or peak shaving, as illustrated in Fig.~\ref{fig:enerflexsim}.

\added{However, existing wholesale market platforms restrict participation to large-scale resources, while prevailing non-market-based demand-response schemes, such as feed-in-tariffs, net metering, and priority dispatch, fail to capture the full system-level potential of \acp{DER}. Moreover, these purely-economic mechanisms neglect the complex preferences inherent to prosumers whose decisions are also shaped by ecological and social values~\cite{georgarakis2021green}, privacy concerns over their proprietary and identifying information~\cite{montakhabi2020sharing}, as well as limited cognitive capacity for market participation~\cite{xia2026bounded}. }

Addressing these limitations, \acp{LEM} are envisioned~\cite{survey_sousa_2019} to facilitate market access for prosumers, including aggregations of residential and commercial buildings, small and medium industries, and energy communities. \Ac{LEM} mechanisms aim to coordinate the operation of \acp{DER} in a prosumer-centric framework, thus unlocking both economic and non-economic value creation. By facilitating local energy exchanges, they have the potential to stimulate investment in DERs, enhance system resilience through reduced dependence on the main grid and provision of flexibility services~\cite{baake2023local}, reduce energy procurement costs and carbon emissions, and enhance community engagement and energy justice~\cite{morstyn2018multiclass}. Emerging platforms such as Beckn\footnote{\url{https://energy.becknprotocol.io/}} and PowerLedger\footnote{\url{https://powerledger.io/solutions/need/p2p/}} exemplify the real-world benefits of \ac{LEM} platforms.\par

\tikzset{
    prosumer/.style={
        draw,
        minimum width=16mm,
        rounded corners,
        fill=blue!10,
        node distance=10mm,
    },
    product/.style={
        draw,
        minimum width=16mm,
        fill=orange!10,
        node distance=10mm,
    },
    arrow/.style={
        -{Latex[length=2mm]},
        line width=0.4mm,
        gray
    }
}
\begin{figure}
    \centering
    \begin{tikzpicture}[line width=0.1mm]
        \node[prosumer] (w) {Wind};
        \node[prosumer, below of=w] (s1) {EV 1};
        \node[prosumer, below of=s1] (d1) {DSO 1};
        \node[product, right of=w, yshift=5mm, node distance=35mm] (e1) {Energy (Hour 1)};
        \node[product, below of=e1] (e2) {Energy (Hour 2)};
        \node[product, below of=e2] (f1) {Flexibility (Hour 1)};
        \node[product, below of=f1] (f2) {Flexibility (Hour 2)};
        \node[prosumer, right of=w, node distance=70mm] (c) {Consumer};
        \node[prosumer, below of=c] (s2) {EV 2};
        \node[prosumer, below of=s2] (d2) {DSO 2};

        \draw[arrow] (w) -- (e1);
        \draw[arrow] (w) -- (e2);
        \draw[arrow] (s1) -- (e1);
        \draw[arrow] (s1) -- (f1);
        \draw[arrow] (s1) -- (f2);
        \draw[arrow] (s2) -- (f1);
        \draw[arrow] (s2) -- (f2);
        \draw[arrow] (s2) -- (e2);

        \draw[arrow] (e2) -- (s1);
        \draw[arrow] (f1) -- (d1);
        \draw[arrow] (f2) -- (d1);
        \draw[arrow] (e1) -- (s2);
        \draw[arrow] (e2) -- (c);
        \draw[arrow] (f1) -- (d2);
        \draw[arrow] (f2) -- (d2);
    \end{tikzpicture}
    \caption{\added{A schematic of the resulting multi-product trade among heterogeneous prosumers in an energy-flexibility market described in Section~\ref{sec:enerflexeg}, that offers energy and flexibility products across two time slots (incoming and outgoing arrows indicate consumption and production, respectively). The prosumers have product interdependent preferences, and the proposed mechanism iteratively finds an equilibrium price for each product using a simple communication format, such that the resulting consumption and production of each product is equal, leading to a successful local exchange.}}\label{fig:enerflexsim}
\end{figure}

\added{A major challenge, limiting the efficiency and attained welfare gains of existing \ac{LEM} mechanisms, is their ability to accurately elicit and clear complex prosumer preferences over heterogeneous and interdependent energy and grid-support products.} The authors in ~\cite{morstyn2018multiclass} and \cite{sorin2018consensus} introduced multi-product \acp{LEM}, with product differentiations accounting for the economic, environmental, and social values of prosumers. \added{These mechanisms increase achieved welfare gains over single-product ones, by unlocking a richer set of mutually beneficial trades across heterogeneous products.} Yet, they fail to capture the interdependencies between products that are inherent to prosumers' preferences, \added{leading to distorted valuations and suboptimal allocations.}
\added{Richer bid formats have been proposed to better capture such interdependencies between products. \cite{Kiedanski_Orda_Kofman_2020} introduced a parametric bid format, while~\cite{bobo2021price} proposed an extended linear non-separable price-region bid format. However, these approaches rely on ad-hoc parameterisations or are limited to specific participant types. Furthermore, eliciting accurate preferences through these bid formats is cognitively prohibitive, as it requires prosumers to analytically express their own preferences over any combination of prices and products~\cite{samuelson2024note}.} Conversely, it is cognitively simpler to determine preferred consumption at given prices. Such a query is routinely faced by market participants, such as under time-of-use pricing. Thus, recent studies have employed \emph{combinatorial} bid formats that allow prosumers to submit a preferred combination of energy products at a given set of prices, that is, a \emph{package}, for multiple price scenarios~\cite{Kiedanski_Kofman_Orda_2021}. This avoids ad-hoc parameterization of prosumer preferences and provides a uniform communication format incorporating diverse and heterogeneous preferences. However, to ensure proper representation of their preferences in market clearing, prosumers must determine preferred packages for numerous price scenarios, which can grow exponentially with the number of products. To mitigate this,~\cite{Hubner_Hug_2025} proposes decision support tools to limit the number of price scenarios needed. Nevertheless, participating in these markets imposes a high cognitive effort on prosumers due to such bid formation tasks.

In contrast, iterative \ac{LEM} clearing mechanisms, where prosumers iteratively report their preferred package of products at announced prices\textemdash referred to as a \emph{package query}, offer prosumers an alternative for accurate yet practical preference elicitation. Instead of submitting a single comprehensive bid encoding complex preferences across all combinations of prices and products, prosumers iteratively report their preferred consumption (or production) of products for an announced set of prices, while the mechanism updates prices based on aggregate responses until convergence. \added{These package queries are inherently intuitive~\cite{samuelson2024note,cramton2006combinatorial}, enabling a more accurate elicitation of combinatorial prosumer preferences, while mitigating the cognitive effort compared to full preference elicitation.
When bidding is automated, the interpretability of package queries further enables prosumers to verify their agent's choices against their own preferences, ensuring transparency without additional cognitive effort.}
Iterative mechanisms have been explored within the context of demand response and load aggregation. The authors in \cite{Tsaousoglou_Steriotis_Efthymiopoulos_Makris_Varvarigos_2020} proposed an iterative mechanism for sourcing demand response services using a \ac{VCG} mechanism. While this non-linear pricing mechanism ensures strategy proofness, its lack of transparency and interpretability raises prosumers' cognitive effort. Alternative approaches in \cite{Chapman_Verbič_2017,Mhanna_Chapman_Verbič_2018} instead adopt an intuitive linear pricing scheme, assigning a unit price to each product, which reduces cognitive effort but remains vulnerable to strategic manipulation. \added{While these approaches reduce the prosumers' cognitive effort per query, convergence may require many query rounds, imposing significant \emph{communication overhead}.} Furthermore, these works assume a single-sided auction architecture with a fixed producer or consumer role, whereas double-sided exchange architectures are more natural and better suited to capturing the full value of peer-to-peer trading in \acp{LEM}, where prosumers act simultaneously as consumers and producers.

\added{Together, the cognitive effort of bid formation tasks and the communication overhead of repeated query rounds constitute what we refer to as the \emph{participation burden}, a critical aspect of market design which can decrease practical usability and lead to suboptimal participation if left unadressed~\cite{scheffel2011experimental}.} Given prosumers' limited cognitive and communication resources, practically-usable \acp{LEM} require transparent pricing and intuitive preference elicitation formats. \added{While the aforementioned \ac{LEM} mechanisms expose clear trade-offs between economic properties and practical usability, none explicitly internalize participation burden as a design objective, either ignoring it or addressing it in an ad-hoc manner through decision support tools~\cite{Hubner_Hug_2025} or simplified parameterisations~\cite{Kiedanski_Orda_Kofman_2020}. Achieving an effective balance between desirable economic properties and practical usability thus remains an open challenge for \ac{LEM} design.}

\added{To address the aforementioned gaps, this paper introduces a unified \ac{LEM} framework that jointly pursues two design objectives rarely considered together: accurate elicitation of combinatorial prosumer preferences, and limited participation burden~\cite{survey_sousa_2019}.} Our contributions are as follows:

\begin{enumerate}
    \item \added{We propose a unified multi-product \ac{LEM} framework, that jointly clears heterogeneous and interdependent energy and grid-support products in a two-sided exchange architecture, extending single-product and separable multi-product designs to a fully combinatorial setting without enforcing a producer-consumer dichotomy;}
    \item \added{We develop a practical and scalable iterative clearing mechanism applied to the proposed \ac{LEM} that combines linear pricing and \textit{intuitive} package queries, to minimize the prosumers' \textit{cognitive effort}, and leverages an ML-aided price discovery algorithm to accelerate convergence and reduce \textit{communication overhead}. Together, these improvements mitigate the prosumers' \textit{participation burden.}}
    \item \added{We conduct a systematic analytical and empirical evaluation of the achievable trade-offs between desirable economic properties, such as incentive compatibility and market efficiency, and practical usability, including participation burden for prosumers and computational complexity for market operators, across diverse \ac{LEM} scenarios.}
\end{enumerate}
\added{Crucially, by treating practical usability as an explicit design objective alongside economic properties, this novel approach equips market designers with the tools to tailor the proposed \ac{LEM} framework to the specific needs and capabilities of their target prosumers, enabling \ac{LEM} designs that are not only theoretically sound but also practically usable.}

\textbf{Outline:} Section~\ref{sec:prelimin} describes the challenges associated with prosumers participating in \acp{LEM} and the proposed framework, while the proposed mechanisms are described in Section~\ref{sec:iterexch}. Section~\ref{sec:numexp} describes the numerical experiments and Section~\ref{sec:conclusion} concludes the paper.

\section{Proposed LEM framework: A multi-product combinatorial exchange}\label{sec:prelimin}

\subsection{Motivation: \acp{LEM} with combinatorial preferences} \label{sec:combexchpros}
\added{We introduce an \ac{LEM} where prosumers can jointly trade multiple energy and grid-support products. Let $\nprod$ be the number of distinct products available in the market, such as electricity or various grid-support services, in different time slots. A prosumer $i$ can trade volume $\product{i,j}{} \in \realN{}$ (negative and positive values represent selling and buying, respectively) of product $j$ at a unit price $\price{j}{} \in \realN{}$. A combination of all products (so-called \textit{package}) is denoted by a vector $\package{i}{}\in\packageset{i}\subseteq\realN{\nprod}$, where the feasible set $\packageset{i}$ is determined by the prosumer's technical constraints.}\footnote{Vectors of products and prices are denoted in corresponding bold symbols.} \added{The value function $\val{i}:\packageset{i}\rightarrow \realN{}$ assigns a numerical value to each feasible package, representing its desirability for prosumer $i$. 
Complexity arises from the combinatorial nature of the prosumers' preferences (value function and feasible set). Because energy and grid-support products are coupled by physical and operational constraints, usage preferences, and socio-ecological concerns, and because prosumers can trade with heterogeneous counterparts or with external entities, creating opportunity costs that further shape their valuations, neither the desirability nor the feasibility of a given trade can in general be decomposed into independent per-product terms. Instead, the value function $\val{i}$ and feasible set $\packageset{i}$ can encode complex complementarities and substitutabilities over all products and are not assumed to be per-product separable}\added{, standing in contrast to more restrictive single-product and separable multi-product \ac{LEM} designs.} The following motivating examples illustrate why such combinatorial preferences naturally arise in practice:

\subsubsection{Motivating example 1: Joint energy-flexibility products}\label{sec:enerflexeg} 

Consider a prosumer $i$ with an \ac{EV} offering energy and flexibility products across two time slots (i.e., $m=4$). \added{Energy products $\product{i,1}{}, \product{i,2}{}$ (in kWh) represent commitments to consume or supply electricity in a time slot, while flexibility products $\product{i,3}{},\product{i,4}{}$ (in kW) represent capacities reserved for upward/downward power adjustments to support grid stability.}
The prosumer's EV is characterized by its \ac{SOE} in each time slot, $\soe{i,1}, \soe{i,2}$, representing the amount of energy available in the \ac{EV}'s battery, where $\soe{i,0}$ denotes the initial \ac{SOE}. \added{The \ac{SOE} evolves dynamically based on the energy charged/discharged:
\begin{equation}
    \soe{i,j} = \soe{i,j-1} + \product{i,j}{}, \quad \forall j \in \{1,2\}. \label{eq:storconssoc1}
\end{equation}
Furthermore, joint feasibility of a combination of energy and flexibility commitments requires that the \ac{SOE} remains within battery capacity $\soecap$ under any activation of upward/downward flexibility commitments, i.e.,
\begin{equation}
    0 \leq \soe{j} \pm \sum_{k=3}^{j+2} \product{i,k}{} \leq \soecap, 
    \quad \forall j \in \{1,2\}. \label{eq:storconsflex1}
\end{equation}
The prosumer's value function reflects two coupled sources of (dis)utility: i) the opportunity cost of trading in the \ac{LEM} rather than with external entities at given buy and sell prices $\linpricehigh{}{}, \linpricelow{}{} \in \mathbb{R}^4$, expressed as $\langle \linpricehigh{}{}, \package{i}{+} \rangle + \langle \linpricelow{}{}, \package{i}{-} \rangle$, with $\package{i}{+}:=\max\{0, \package{i}{}\}$ and $\package{i}{-}:=\min\{0, \package{i}{}\}$ denoting the positive and negative parts of $\package{i}{}$; and ii) the discomfort incurred for deviating from the desired \ac{SOE} $\soer{j}$ over a sequence of time slots. The instantaneous discomfort incurred can typically be expressed as an L2-norm $ \rVert{} \soe{} - \soer{} \rVert_{\soepenalty{}}^2$, with penalty weights $\soepenalty{}$. In this setting, the non-separability of the resulting value function $\val{i}(.)$ of this prosumer is a direct consequence of the intertemporal \ac{SOE} coupling and joint feasibility constraints link energy and flexibility products across time. Indeed, the value of offering flexibility in slot $j$ depends on the \ac{SOE} at that slot, which is itself shaped by the energy traded in all preceding slots, effectively coupling the valuation of a single product at each time slot to the full sequence of packages traded.}

For announced linear prices $\linprice{}{} \in \mathbb{R}^{4}$, this prosumer can efficiently compute its optimal package allocation by solving a constrained utility-maximization problem, \added{internalizing all cross-product and inter-temporal couplings without need for simplifying assumptions}:
\begin{subequations} \label{eq:EV_max}
    \begin{align}
        \max_{\package{i}{}, \mathbf{s}}\ & \left(\langle \linpricehigh{}{}, \package{i}{+} \rangle + \langle \linpricelow{}{}, \package{i}{-} \rangle - \lVert \soe - \soer \rVert_{\soepenalty{}}\right) - \langle \linprice{}{}, \package{i}{} \rangle \label{eq:storobj}\\
        \text{s.t.}\ & \text{Eqs.} \eqref{eq:storconssoc1}-\eqref{eq:storconsflex1} 
    \end{align}
\end{subequations}
\added{The difficulty arises when a separable bid format requires the prosumer to decompose this valuation structure into independent per-product and per-time demand curves, either by assuming fixed price forecasts to approximate the standalone marginal values of coupled products, distorting true preferences, or by adopting conservative feasibility sets that exclude valuable jointly feasible trades. In contrast, a multi-product \ac{LEM} design that directly clears packages jointly with full combinatorial preference elicitation avoids these losses, internalizing all complementarities, and enabling higher-value allocations.}

\subsubsection{Motivating example 2: Socio-ecological product differentiation} \label{sec:social-eco-prefs}

In addition to the previous motivating example, \acp{LEM} can allow prosumers to trade multiple differentiated energy labels, representing their technical, social, and ecological concerns, for instance, inspired by \cite{morstyn2018multiclass}: \emph{Standard} (non-renewable grid imports), \emph{Green} (renewable production, regardless of geographical origin), and \added{\emph{Local} (community-based production, regardless of energy source)}. \added{As these labels are constrained \emph{substitutes}, the marginal value of each label inherently depends on the full composition of the package. Beyond this structural coupling, different prosumer types introduce further combinatorial features, for instance:}
\added{\emph{Eco-conscious} and \emph{community-minded} prosumers may value not only the total quantity of green energy consumed, but its \emph{share} in their total consumption: $\frac{\product{i,\text{G}}{+}}{\sum_{k\in \{\text{St},\text{G},\text{L}\}}\product{i,k}{+}}$, and attach additional value to a balanced consumption supporting both green and local sources, creating a supermodular complementarity between these labels: $ \mu_i \product{i,\text{L}}{+} + \nu_i \product{i,\text{G}}{+} + \rho_i \min\!\left(\product{i,\text{L}}{+},\product{i,\text{G}}{+}\right),$ where $\mu_i, \nu_i \geq 0$ are standalone label value, and $\rho_i > 0$ is a balance premium. Their valuation of green and local energy depends, therefore, on the composition of the entire package consumed, and no pair of prices $(\price{\text{L}}{}, \price{\text{G}}{})$ can accurately reflect the standalone values of these labels.}

\added{These examples reveal a fundamental limitation of existing \ac{LEM} mechanisms, which restrict and distort preference elicitation through single-product or separable multi-product bid formats. Addressing this limitation is the central challenge addressed in this paper, seeking to design a bid format that is both \textit{expressive} enough to faithfully represent combinatorial prosumer preferences and \textit{intuitive} enough for practical use.}

\subsection{Combinatorial exchange market clearing}
We introduce the following combinatorial exchange \ac{LEM} with $\npros$ prosumers. Each prosumer $i$ is allocated a package of products $\package{i}{}$. Based on the prosumers' value functions $\val{i}:\packageset{i}\rightarrow \realN{}$ and feasible sets $\packageset{i}$, the market operator seeks to maximize social welfare as
\begin{subequations}\label{eq:marketmodel}
    \begin{align}
        p^\star = \max_{\{\package{i}{}\in\packageset{i}\}_{i=1}^\npros} \ & \sum_{i=1}^{\npros}\val{i}(\package{i}{})\label{eq:marketobjective}\\
    \text{s.t.} \ & \sum_{i=1}^{\npros} \package{i}{} = \mathbf{0},\label{eq:marketobjtradebal}
    \end{align}
\end{subequations}
where~\eqref{eq:marketobjtradebal} enforces trade balance for each product.

\subsection{Linear pricing rule and Walrasian equilibrium}\label{sec:walrasian}

Under the proposed linear pricing rule, each product $j \in \mathcal{M}$ is associated with a unit price $\linprice{j}{}$, and the total payment for a package $\package{i}{}$ is $\langle \package{i}{}, \linprice{}{}\rangle$. Under the assumption of concavity of the value functions and mild assumptions on the feasible sets, strong duality holds for \eqref{eq:marketmodel}, and these linear market clearing prices $\linprice{}{\star}$ are obtained as a solution of the corresponding dual program
\begin{equation}\label{eq:lagrangedual}
    g^\star = \min_{\linprice{}{}} \lagrangedual(\linprice{}{}) :=  \sum_{i=1}^{\npros}\, \max_{\package{i}{}\in\packageset{i}} \left(\val{i}(\package{i}{}) - \langle\linprice{}{}, \package{i}{}\rangle\right)
\end{equation}
such that they achieve a Walrasian equilibrium~\cite[Chapter~5.6]{boyd2004convex}. Formally, this means that the aggregate preferred packages of all prosumers at prices $\linprice{}{\star}$ are balanced for each product, i.e. $\sum_{i=1}^\npros \package{i}{\star}(\linprice{}{\star}) = \mathbf{0}$, where 
\begin{equation} \label{eq:best_response}
    \package{i}{\star}(\linprice{}{\star}) \in \arg\max_{\package{i}{}\in\packageset{i}}\ \left(\val{i}(\package{i}{}) - \langle\linprice{}{\star}, \package{i}{}\rangle\right)
\end{equation} 
denotes $i^\text{th}$ prosumer's preferred package at prices $\linprice{}{\star}$.

Yet, in practically-relevant applications, non-concave value functions commonly arise from integral constraints associated with flexible assets such as batteries and switchable devices, or combinatorial preferences over products inherent to prosumers. In these cases, strong duality fails, and thus, the existence of linear clearing prices that achieve a Walrasian equilibrium is not guaranteed. As a result, for prices $\linprice{}{\star}$, which are a solution of the dual program~\eqref{eq:lagrangedual}, some products may exhibit unmet demand or excess supply\textemdash \emph{imbalances}.

While non-linear pricing can support equilibrium in non-concave economies by making unit prices package-dependent, linear prices are preferred in practice, as they are intuitive and transparent, thus encouraging broader participation in \acp{LEM}. In this paper, we bridge the gap between theoretical market properties and practical applications by showing that, in large, real-world settings with many prosumers, linear prices approximately clear, residual imbalances are small, and welfare losses are negligible.

\subsection{Duality gap for non-concave prosumers}\label{sec:dualgapshapley}

\added{Authors in~\cite{udell2016bounding,hreisson2021folkman} analyzed the duality gap in optimization problems involving a sum of non-concave functions subject to linear equality and inequality constraints, deriving from the Shapley-Folkman-Starr theorem~\cite{starr1969quasi}. The key idea is to bound the non-concavity of the sum of functions by that of one of the functions, thereby bounding the duality gap. We now introduce a few useful definitions.}

\begin{definition}[Concave envelope of a function]
    \added{Concave envelope $\overline{v}_i: \packageset{i} \rightarrow \mathbb{R}$ of a function $\val{i}: \packageset{i} \rightarrow \mathbb{R}$ is the smallest concave function minorised by $\val{i}$, i.e., $\val{i}(\package{i}{}) \leq \overline{v}_i(\package{i}{}),\ \forall \package{i}{}\in\packageset{i}$.}
\end{definition}

\begin{definition}[Non-concavity of a function]
    \added{Non-concavity of a function $\val{i}$ is defined as}
    \begin{equation*}
        \operatorname{ncvx}(\val{i}) = \max_{\package{i}{}\in\packageset{i}}\, \overline{v}_i(\package{i}{}) - \val{i}(\package{i}{})
    \end{equation*}
\end{definition}

\begin{theorem}
    \added{For optimization problem~\eqref{eq:marketmodel}, where the objective function is a sum of non-concave functions subject to linear equality and inequality constraints, the duality gap is less than or equal to the largest non-concavity~\cite{udell2016bounding}, i.e.,}
    \begin{equation*}
        g^\star - p^\star \leq \max_{i\in\intset{1}{\npros}} \operatorname{ncvx}(\val{i})
    \end{equation*}
\end{theorem}

Thus, with an increase in prosumers, $p^\star$ increases, and the relative duality gap $\rho^\text{dual}:=(g^\star - p^\star)/p^\star$ vanishes, making problem~\eqref{eq:marketmodel} approximately concave. Based on these results, we establish the validity of linear pricing in \acp{LEM} with a large number of heterogeneous prosumers and the supporting numerical evidence is presented in Section~\ref{sec:numexp}.

\section{Intuitive and scalable preference elicitation}\label{sec:iterexch}


As illustrated in prior motivating examples, it is rarely practical for prosumers with combinatorial preferences to evaluate their value over all product bundles. For instance, the EV's value function in Example~\ref{sec:enerflexeg} is implicitly defined by~\eqref{eq:EV_max}; results in non-linear, non-separable value function incompatible with predetermined bid formats, lacking a closed-form expression and intuitive per-product or per-hour interpretation.

\subsection{\Ac{CCE}}\label{sec:cce}

To tackle these challenges, we introduce a Combinatorial Clock Exchange (CCE) mechanism for \acp{LEM}. This is an iterative procedure, where at each iteration, the market operator announces product prices and elicits package queries from prosumers. In response, each prosumer reports their preferred package by solving their utility-maximization problem~\eqref{eq:best_response} for the given linear prices. The aggregate response of prosumers to these queries may reveal shortages or surpluses for each product, based on which prices are adjusted before the next iteration. The term \emph{clock} refers to the repeated price adjustment for each product, while \emph{exchange} refers to the double-sided market, where the market operator facilitates trading among prosumers without acting as a buyer or seller.
However, determining suitable price adjustments is non-trivial due to prosumer-specific product interdependencies\textemdash different products may act as substitutes or complements for each prosumer, making the allocation problem \emph{combinatorial}. We note that the Lagrange dual function $\lagrangedual$ defined in~\eqref{eq:lagrangedual} is convex in $\linprice{}{}$, as it is the pointwise maximum of affine functions, regardless of whether the underlying value functions are concave, whose gradient is given by
\begin{equation}
    \nabla \lagrangedual(\linprice{}{}) = -\sum_{i=1}^{\npros} \package{i}{\star}(\linprice{}{}),
\end{equation}
which is conveniently an aggregate of the elicited package queries. Then, the prices for the next iteration are obtained as
\begin{equation}
    \linprice{}{t+1} = \linprice{}{t} - \lrcce{t} \nabla\lagrangedual(\linprice{}{t}),
\end{equation}
where the price update direction aligns with the imbalances, increasing prices for products with shortages and decreasing them for surpluses, gradually converging to a solution of~\eqref{eq:lagrangedual}. The procedure is guaranteed to converge if the time-dependent step sizes $\lrcce{t}$ satisfy the Robbins-Monro conditions for non-smooth optimization~\cite{garrigos2023handbook}, specifically, ${\sum_t \lrcce{t} = \infty,\ \sum_t \lrcce{t}^2 < \infty}$.

\subsection{\Ac{MLCCE}}
In the proposed \ac{CCE}, we use only the latest elicited package queries and pre-determined step sizes for price update at each iteration. To facilitate faster convergence, we introduce an ML-aided price update method, which utilizes all the previous package queries to learn prosumers' preferences through \ac{IO}~\cite{chan2025inverse}, and improve price updates.

\subsubsection{Value function estimation}
Let the set of received package queries from prosumer $i$ until iteration $t$ be $\mathcal{D}^t_i:= \{(\dpackage{i}{k}, \linprice{}{k})\}_{k=1}^t$. In \ac{IO}, the objective is to learn the prosumer's value function using a parameterized model $\lval{i}:\packageset{i}\rightarrow\mathbb{R}$ with parameters $\params{}{}$, such that the estimated preferred packages
\begin{equation}\label{eq:lprosumerproblem}
    \lpackage{i}{k} \in \arg\max_{\package{i}{}\in\packageset{i}}\, \hat{v}_i(\package{i}{}; \theta) - \langle\linprice{}{k}, \package{i}{}\rangle
\end{equation}
align with the prosumer's true preferred packages $\dpackage{i}{k}$. In order to determine appropriate parameters $\params{}{}$, we minimize the suboptimality loss~\cite{chan2025inverse} as
\begin{subequations}\label{eq:invopt}
    \begin{align}
        \min_{\params{}{}} \ & \sum_{k=1}^t \lval{i}(\lpackage{i}{k}; \params{}{}) - \langle\linprice{}{k}, \lpackage{i}{}\rangle - [\lval{i}(\dpackage{i}{k}; \params{}{}) - \langle\linprice{}{k}, \dpackage{i}{}\rangle] \label{eq:invoptloss}\\
        \text{s.t.}\ & \lpackage{i}{k} \in \arg\max_{\package{i}{}\in\packageset{i}}\, \hat{v}_i(\package{i}{}; \theta) - \langle\linprice{}{k}, \package{i}{}\rangle, \, \forall k\in\intset{1}{t},\label{eq:invoptoptpack}
    \end{align}
\end{subequations}
where~\eqref{eq:invoptloss} defines the suboptimality loss\textemdash the difference between the estimated utility at the estimated and true preferred package. It is non-negative by construction, as $\lpackage{i}{k}$ is a maximizer, and becomes zero when the estimated and true preferred packages coincide for all $t$ queried prices. Thus, minimizing suboptimality loss ensures that the estimated value function captures the true preferences.

For the estimator $\lval{}$, we employ a class of monotone parametric functions, as prosumer's value is monotone in products. Specifically, we adopt the \acp{MVNN} introduced in~\cite{Weissteiner_Heiss_Siems_Seuken_2022}, modified for prosumers to accommodate both consumption and production, by extending their domain and range to negative values. Since \acp{MVNN} use bounded-ReLU activations, the optimization problem in~\eqref{eq:invoptoptpack} can be formulated as a \ac{MILP}~\cite{Fischetti_Jo_2018}. The detailed formulation and the procedure for solving~\eqref{eq:invopt} are provided in Appendix~\ref{app:mvnnprosumerprob} and Algorithm~\ref{alg:io}, respectively, where we solve~\eqref{eq:invopt} iteratively using standard neural network gradient descent optimizers. Specifically, we compute the estimated preferred packages in~\eqref{eq:invoptoptpack} for the current network parameters, to compute loss~\eqref{eq:invoptloss} at each gradient descent step. We next describe how the estimated value function is used to generate enhanced price iterates.

\subsubsection{Price update}
In \ac{CCE}, the step sizes were pre-determined. Here, we compute suitable step sizes by minimizing the estimated Lagrange dual function $\llagrangedual(\linprice{}{}) = \sum_{i=1}^\npros\, \max_{\package{i}{}\in\packageset{i}} (\lval{}(\package{i}{}) - \langle \linprice{}{}, \package{i}{}\rangle)$ as
\begin{subequations}\label{eq:mlcceupdate}
    \begin{align}
        \lrmlcce{t} \in \arg\min_{\lrcce{}}\quad & \hat{g}(\linprice{}{t} - \lrcce{}\nabla g(\linprice{}{t}))\\
        & \lrcce{} \in [\underline{\eta}/\sqrt{t}, \bar{\eta}/t], \label{eq:robbinsconst}
    \end{align}
\end{subequations}
where $\underline{\eta}$, $\bar{\eta}$ denotes positive constants. \eqref{eq:robbinsconst} confines the step size between two sequences that satisfy the Robbins-Monro conditions, thereby so does the resulting sequence $\lrmlcce{t}$. Problem~\eqref{eq:mlcceupdate} is convex in $\lrcce{}$, since $\llagrangedual$ is convex as a pointwise maximum of affine functions, and convexity is preserved under affine composition. Thus, we solve it using gradient descent until the gradient vanishes or a boundary point in~\eqref{eq:robbinsconst} is reached. 

\added{The step size calculated above is used for the price update, at each iteration. The resulting set of prices are announced in the next iteration and package queries are elicited. After each round of package queries, an imbalance index is evaluated. Let $\Delta^\text{imb}(\package{}{}) := \sum_{i=1}^\npros \package{i}{}$ denote trade imbalance for an allocation $\package{}{}:=\{\package{i}{}\}_{i=1}^\npros$, then the imbalance index is defined as}
\begin{equation}
   \rho^\text{imb}(\package{}{}) := \frac{1}{m}\sum_{j=1}^\nprod \lvert\Delta^\text{imb}( \package{}{})_j\rvert / \maxpackage{j},
\end{equation}
\added{where $\maxpackage{j}:=\sum_{i=1}^\npros \max_{\package{i}{}\in\packageset{i}}\,\lvert\product{i,j}{}\rvert$ represents the maximum feasible imbalance volume for product $j$.}

Finally, due to an initial lack of query data for value estimation, the \ac{CCE} price update is run for the first $t^0$ iterations. The complete \ac{MLCCE} procedure is described in Algorithm~\ref{alg:mlcce}.

\begin{algorithm}[t]
    \caption{ML-aided combinatorial clock exchange}\label{alg:mlcce}
    \begin{algorithmic}[1]
        \State Input: $\nprod$, $\npros$, $\linpricelow{}{}$, $\linpricehigh{}{}$, $\{\lrcce{t}\}_{t=0}^{t^0-1}$, $\{\packageset{i}\}_{i=1}^\npros$, $\horizon$, $t^0$
        \State Init: Prices $\linprice{j}{0} \leftarrow (\linpricelow{j}{} + \linpricehigh{j}{})/2,\, \forall j\in\intset{1}{\nprod}$, ${\dataset{i}{} \leftarrow \{\}}$, $t^\star \leftarrow 0$, $\rho^{\text{imb}\star}\leftarrow \mathbf{1}$
        \Procedure{Clock phase}{}
            \For{$t\leftarrow \intset{0}{T-1}$}
                \State Announce $\linprice{}{t}$
                \State Query $\dpackage{i}{t},\, \forall i\in\intset{1}{\npros}$
                \If{$\rho^\text{imb}(\dpackage{}{t}) < \rho^{\text{imb}\star}$} (save best iterate) 
                    \State $\rho^{\text{imb}\star}\leftarrow \rho^\text{imb}(\dpackage{}{t}),\ t^\star\leftarrow t$
                \EndIf
                \If{$t<t_0$} (\ac{CCE} price update)
                    \State $\linprice{}{t+1} \leftarrow \linprice{}{t} + \lrcce{t}\sum_{i=1}^\npros \dpackage{i}{t}$
                    
                \Else{} (\ac{MLCCE} price update)
                    \State Update data:
                    \State \hfill $\dataset{i}{} \leftarrow \dataset{i}{} \cup (\dpackage{i}{t}, \linprice{}{t}),\ \forall i\in\intset{1}{\npros}$
                    \State Train $\{\lval{i}\}_{i=1}^n$ using Algorithm~\ref{alg:io}
                    \State Compute $\linprice{}{t+1} \gets$ solution of Problem~\eqref{eq:mlcceupdate} 
                \EndIf
            \EndFor
        \EndProcedure
        \State \Return Best iteration: $t^\star$
    \end{algorithmic}
\end{algorithm}

\subsection{Feasible allocation}\label{sec:feasallo}
The above procedures determine the clearing price $\linprice{}{\star}$ as solutions of~\eqref{eq:lagrangedual}. The resulting dispatch $\package{i}{\star}(\linprice{}{\star}),\, \forall i\in\intset{1}{\npros}$ may exhibit small imbalances in the absence of strong duality, as discussed in Section~\ref{sec:walrasian}. To restore trade balance, the market operator procures and sells the excess demand and supply at external prices $\linpricehigh{}{}$ and $\linpricelow{}{}$, respectively, incurring a revenue deficit for each product $j$, given by
\begin{equation*}
     \Delta^{\text{rev}}_j = (\linpricehigh{j}{} - \linprice{j}{\star})\left( \sum_{i=1}^\npros \product{i,j}{} \right)^+ + (\linpricelow{j}{} - \linprice{j}{\star})\left( \sum_{i=1}^\npros \product{i,j}{} \right)^- ,
\end{equation*}
representing the cost of implementing a linear price rule that facilitates market participation for prosumers with limited cognitive capacity. The deficit can be socialized via a unit trade fee, which for a product $j$ is $\rho^\text{rev}_j = \Delta^\text{rev}_j / \sum_{i=1}^\npros \lvert \product{i,j}{} \rvert $, where the denominator denotes the total traded volume of product $j$. This unit trade fee becomes negligible in large markets due to the vanishing relative duality gap $\rho^\text{dual}$ and imbalance.

\subsection{Desirable mechanism properties}\label{sec:irvsbb}

The proposed mechanism does not fully incentivize truthful reporting, since linear pricing allows prosumers to potentially benefit by misreporting their preferred package $\package{i}{\star}(\linprice{}{})$ for announced prices $\linprice{}{}$. However, these benefits diminish with increasing market size and decreasing market power, making this mechanism approximately incentive compatible ~\cite[Chapter 10]{cramton2006combinatorial}. Further, due to the diminishing relative duality gap in large markets, the mechanism finds welfare optimal allocation and thus, is approximately efficient.

The revenue deficit, which is socialized via the post clearing trade fee $\rho^\text{rev}$, makes the proposed mechanism budget balanced for the market operator (i.e., no missing money problem). Further, the clock phase is individually rational, since it allocates the prosumers' preferred packages for posted prices, ensuring that prosumers have non-negative utilities. However, due to the post clearing trade fee, prosumers might experience negative utility. Again, due to the diminishing unit trade fee, the mechanism is approximately individually rational for large markets.

\section{Numerical experiments}\label{sec:numexp}
We now validate the above claims and evaluate the proposed mechanisms on several market instances. A market instance is characterized by the participating prosumers' preferences and tradeable products, where we generate prosumer preferences stochastically. In addition to the \ac{EV} described in Example~\ref{sec:enerflexeg}, other prosumer types are detailed in the online appendix in the Github repository~\cite{shobhit_singhal_2025_17241773}. The following figures show the average over 10 randomly generated instances, with shaded region indicating the standard deviation. All the experiments are implemented in Python 3.13 and run on an Apple M3 8-core CPU with 16GB of RAM.

\subsection{Social welfare in multi-product vs product-specific markets}\label{sec:resjointmar}

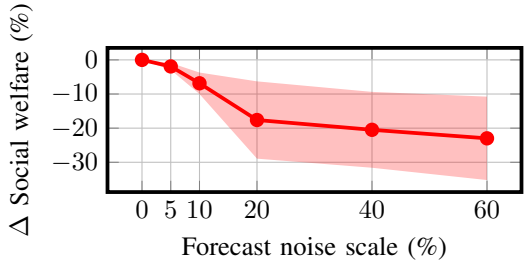
\begin{figure}[t]
    \centering
    \begin{tikzpicture}
        \begin{axis}[
            height=3.5cm,
            width=0.8\linewidth,
            table/col sep=comma,
            axis line style={line width=0.5mm},
            xtick = {0,5,10,20,40,60},
            grid=major,
            xlabel = Forecast noise scale (\%),
            ylabel = $\Delta$ Social welfare (\%)
        ]
            \addplot[line width=0.5mm, red, mark=*]
                table[y=Mean, x index=0] {seq_joint.csv};
            \addplot[opacity=0, name path = upper] 
                table[y=Upper, x index=0] {seq_joint.csv};
            \addplot[opacity=0, name path = lower] 
                table[y=Lower, x index=0] {seq_joint.csv};
            \addplot[fill=red, fill opacity=0.25] 
                fill between[of = upper and lower];
        \end{axis}
    \end{tikzpicture}
    \caption{Change in social welfare of sequential relative to the joint market as a function of flexibility price forecast errors.}\label{fig:socwelseqjoint}
\end{figure}
Here, we simulate the \ac{LEM} described in Example~\ref{sec:enerflexeg}, considering two market structures: a joint energy-flexibility market and a sequential market, where the energy market is cleared first, followed by the flexibility market. The \ac{LEM} involves six participants and the resulting trade for the joint market is shown in Fig.~\ref{fig:enerflexsim}. In the sequential case, prosumers must forecast flexibility prices to bid in the energy-only market, breaking energy-flexibility complementarities.  The detailed prosumer's problem is described in the online appendix~\cite{shobhit_singhal_2025_17241773}. Prosumers must hedge against forecast error, either withholding unnecessary \ac{SOE} headroom or under-committing flexibility. We observe that increasing levels of forecast error reduces achieved social welfare, as shown in Fig.~\ref{fig:socwelseqjoint}. Not only do product-specific market structures impose a significant cognitive burden on prosumers through tasks like price forecasting, but they also limit the expression of product interdependencies, thereby limiting achievable social welfare.

\subsection{Diminishing duality gap and imbalance in large markets}\label{sec:residualgap}
\begin{figure}
    \centering
    \begin{tikzpicture}
        \begin{semilogyaxis}[
            axis line style={line width=0.5mm},
            line width=0.5mm,
            table/col sep=comma,
            xtick = {10, 20, 40, 80, 120},
            ytick = {1, 0.1, 0.01},
            width = 0.8\linewidth,
            height = 4cm,
            xlabel = Market size,
            ylabel = $\rho^\text{dual}$ (\%),
            legend style = {line width=0.25mm, xshift=-1.5cm, yshift=0.02cm, draw=none},
            ymax = 10,
            ymin=0.001,
            grid=major,
        ]
            \addplot[line width=0.5mm, red, mark=*] table[y=Mean, x index=0] {dual_gap_data.csv};
            \addlegendentry{$\rho^\text{dual}$}

            \addplot[name path = up, red, opacity=0] table[x index=0, y=Upper] {dual_gap_data.csv};
            \addplot[name path = lo, red, opacity=0] table[x index=0, y=Lower] {dual_gap_data.csv};

            \addplot[color=red, fill opacity=0.25, fill=red] fill between[of = up and lo];
        \end{semilogyaxis}
        \begin{semilogyaxis}[
            line width=0.5mm,
            table/col sep=comma,
            xticklabels = {},
            ytick = {1, 0.1, 0.01},
            xtick = {10, 20, 40, 80, 120},
            width = 0.8\linewidth,
            height = 4cm,
            ylabel = $\rho^\text{imb}$ (\%),
            legend style = {line width=0.25mm, draw=none, yshift=0.02cm},
            ymax = 10,
            ymin=0.001,
            axis y line* = right
        ]
            \addplot[line width=0.5mm, blue, mark=*] table[y=Mean, x index=0] {imb_norm.csv};
            \addlegendentry{$\rho^\text{imb}$}

            \addplot[name path = up, blue, opacity=0] table[x index=0, y=Upper] {imb_norm.csv};
            \addplot[name path = lo, blue, opacity=0] table[x index=0, y=Lower] {imb_norm.csv};

            \addplot[color=blue, fill opacity=0.25, fill=blue] fill between[of = up and lo];
        \end{semilogyaxis}
    \end{tikzpicture}
    \caption{Duality gap and imbalance as a function of market size (number of prosumers) for a market with 24 hourly energy products.}\label{fig:dualgapmarketsize}
\end{figure}
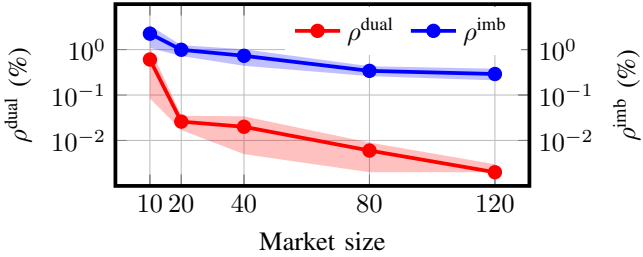

Here, we simulate an \ac{LEM} instance with 24 hourly energy products and varying number of prosumers involving batteries, heat pumps, wind power producers, consumers, and switchable devices. \added{Then, Fig.~\ref{fig:dualgapmarketsize} shows the diminishing relative duality gap and imbalance index with increasing market size which validates the applicability of linear pricing, and provides an empirical evidence for market properties discussed in Section~\ref{sec:irvsbb}.}

\subsection{\added{Convergence rate and computational complexity}}
\begin{figure}[t]
  \centering
  \begin{tikzpicture}[line width=0.5mm]
    \begin{groupplot}[
            group style = {
                group size=1 by 2,
                vertical sep = 0
            },
            width=0.9\linewidth,
            height=4cm
        ]
        \nextgroupplot[
            axis line style={line width=0.5mm},
            table/col sep=comma,
            xtick = {0, 10, 20, 30, 40},
            ylabel = $\rho^\text{imb}$ (\%),
            legend style = {xshift=0, line width=0.25mm},
            grid=major,
            xticklabels={},
        ]
            \addplot[line width=0.5mm, blue] table[y=mean_mlcce, x index=0] {random24_120.csv};
            \addlegendentry{MLCCE}
            \addplot[line width=0.5mm, red] table[y=mean_cce, x index=0] {random24_120.csv};
            \addlegendentry{CCE}
            
            \addplot[name path = cce_up, red, opacity=0] table[x index=0, y=cce_upper] {random24_120.csv};
            \addplot[name path = cce_lo, red, opacity=0] table[x index=0, y=cce_lower] {random24_120.csv};

            \addplot[name path = mlcce_up, blue, opacity=0] table[x index=0, y=mlcce_upper] {random24_120.csv};
            \addplot[name path = mlcce_lo, blue, opacity=0] table[x index=0, y=mlcce_lower] {random24_120.csv};

            \addplot[color=red, fill opacity=0.25, fill=red] fill between[of = cce_up and cce_lo, soft clip={domain=0:40}];
            \addplot[color=blue, fill opacity=0.25, fill=blue] fill between[of = mlcce_up and mlcce_lo, soft clip={domain=0:40}];
            \node[text width=1.3cm, align=center, fill=white] at (axis cs:35,11) {$\nprod=24$\\$\npros=120$};
        \nextgroupplot[
            axis line style={line width=0.5mm},
            table/col sep=comma,
            xtick = {0, 10, 20, 30, 40},
            xlabel = Clock iteration,
            ylabel = $\rho^\text{imb}$ (\%),
            grid=major,
        ]
            \addplot[line width=0.5mm, blue] table[y=mean_mlcce, x index=0] {random6_40.csv};
            \addplot[line width=0.5mm, red] table[y=mean_cce, x index=0] {random6_40.csv};
            
            \addplot[name path = cce_up, red, opacity=0] table[x index=0, y=cce_upper] {random6_40.csv};
            \addplot[name path = cce_lo, red, opacity=0] table[x index=0, y=cce_lower] {random6_40.csv};

            \addplot[name path = mlcce_up, blue, opacity=0] table[x index=0, y=mlcce_upper] {random6_40.csv};
            \addplot[name path = mlcce_lo, blue, opacity=0] table[x index=0, y=mlcce_lower] {random6_40.csv};

            \addplot[color=red, fill opacity=0.25, fill=red] fill between[of = cce_up and cce_lo, soft clip={domain=0:40}];
            \addplot[color=blue, fill opacity=0.25, fill=blue] fill between[of = mlcce_up and mlcce_lo, soft clip={domain=0:40}];

            \node[text width=1.1cm, align=center] at (axis cs:35,7.5) {$\nprod=6$\\$\npros=40$};
    \end{groupplot}
  \end{tikzpicture}
  \caption{Imbalance as a function of clock iterations for \ac{CCE} and \ac{MLCCE}, in a market with 24 hourly energy products and 120 prosumers (top), and 6 hourly products with 40 prosumers (bottom).}\label{fig:resccemlcce}
\end{figure}
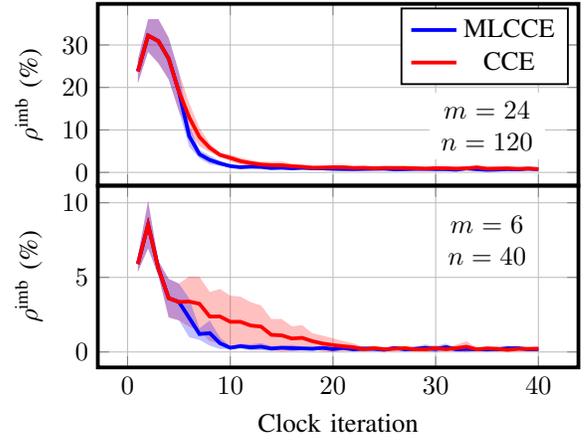


Fig.~\ref{fig:resccemlcce} shows the evolution of imbalance for \ac{CCE} and \ac{MLCCE} in two market instances. The step sizes for \ac{CCE} were pre-tuned for each market instance, resulting in $\lrcce{t}=0.15/t^{0.6}$ (top) and $\lrcce{t}=0.5/t^{0.6}$ (bottom). In contrast, \ac{MLCCE} automatically finds suitable step sizes without manual tuning for each instance, and achieves convergence in fewer iterations. However, these improvements yield increased computational complexity for market operators compared to the \ac{CCE}, requiring $200$ (top) and $20$ (bottom) seconds per clock iteration. Most of the computational effort was spent on learning the value function, where the main bottleneck arose from repeatedly computing the estimated preferred packages in~\eqref{eq:invoptoptpack}, which requires solving the \ac{MILP} problem~\eqref{eq:mvnnmilp}.
\added{Thus, the computational complexity for \ac{MLCCE} is directly proportional to the number of participants.}

\added{\ac{MLCCE} is particularly useful in instances where the cost of conducting a round of package query is expensive. In practical settings, where the market is conducted repeatedly, and assuming that participant preferences do not change significantly, using trained value models from previous market instances can significantly reduce the number of iterations.}

\section{Conclusion}\label{sec:conclusion}
In this article, we designed a combinatorial clock exchange for local energy markets that iteratively elicits prosumer preferences through package queries at announced prices and implements linear pricing. Linear pricing and package queries make participation transparent and accessible for prosumers, fostering trust while maintaining allocative efficiency. To accelerate convergence, we introduced an ML-aided price discovery method that adaptively determines step sizes without domain-specific tuning. Finally, numerical experiments demonstrate benefits of a multi-product over product-specific market structure, validate the practicality of linear pricing, and demonstrate the effectiveness of the proposed mechanisms.

\section*{Acknowledgement}
\added{The authors thank Thomas Falconer (DTU) for valuable comments that helped improve the quality of this paper.}

\section*{Declaration of generative AI and AI-assisted technologies in the writing process}
During the preparation of this work the authors used Llama3.2 in order to improve the readability of certain parts of the manuscript. After using this tool, the authors reviewed and edited the content as needed and take full responsibility for the content of the publication.

\renewcommand{\bibfont}{\footnotesize}
\printbibliography

{
\appendices

\section{Estimating preferred packages with \ac{MVNN}}\label{app:mvnnprosumerprob}
Let $K$ denote the hidden layers in \ac{MVNN} $\lval{}:\packageset{}\rightarrow\mathbb{R}$, where each layer $k\in\intset{1}{K}$ has $n_k$ units and is composed with a bReLU activation unit. Let layer $0$ denote input layer with $\nprod$ units and $K+1$ denote the output layer with a single unit. Let $\mathbf{w}_{k,j}, b_{k,j}$ denote the weights and biases for unit $j$ and $k$, and $\{0,t_{k,j}\}$ denote the saturation limits for the activation unit. These network parameters are grouped in $\params{}{}:=\{\mathbf{w}, \boldsymbol{b}, \boldsymbol{t}\}$. Let $\soutput{k}{j}$ denote the output of unit $j$ in layer $k$, where $\voutput{0}$ denotes the input of length $\nprod$ and $\soutput{K+1}{0}$ denotes the final output. The input and output layers are not followed by an activation unit. Then, the training procedure to solve~\eqref{eq:invopt} is described in Algorithm~\ref{alg:io}, where the prosumer's utility maximization problem~\eqref{eq:lprosumerproblem} with value estimate $\lval{}$ reads as
\begin{subequations}\label{eq:mvnnmilp}
    \begin{align}
        y_0^\star(\linprice{}{}; \params{}{}) \in \arg & \max_{\Psi} \  \soutput{K+1}{0} - \langle \linprice{}{}, \voutput{0}\rangle\\
        \text{s.t.} \ & \voutput{0} \in \packageset{} \label{eq:mvnnmilp_packagecons}\\
        & s_{k,j} = \langle\mathbf{w}_{k,j}, \voutput{k-1}\rangle + b_{k,j},\nonumber\\ 
        & \hspace{0.6cm} \forall k\in\intset{1}{K}, j\in\intset{1}{n_k}\\
        & \soutput{k}{j} \leq \alpha_{k,j} t_{k,j},\nonumber \\
        & \hspace{0.6cm} \forall k\in\intset{1}{K}, j\in\intset{1}{n_k}\\
        & \soutput{k}{j} \leq s_{k,j} + M(1-\alpha_{k,j}),\nonumber \\
        & \hspace{0.6cm} \forall k\in\intset{1}{K}, j\in\intset{1}{n_k}\\
        & \soutput{k}{j} \geq \beta_{k,j} t_{k,j},\nonumber \\
        & \hspace{0.6cm} \forall k\in\intset{1}{K}, j\in\intset{1}{n_k}\\
        & \soutput{k}{j} \geq s_{k,j} + (t_{k,j} - M)\beta_{k,j},\nonumber \\
        & \hspace{0.6cm} \forall k\in\intset{1}{K}, j\in\intset{1}{n_k}\\
        & \soutput{K+1}{0} = \langle\mathbf{w}_{K+1,0}, \voutput{K}\rangle + b_{K+1,0}\\
        & \alpha_{k,j}, \beta_{k,j}\in\{0, 1\}
    \end{align}
\end{subequations}
where $\Psi:=\{ \voutput{0},\dots,\voutput{K},\soutput{K+1}{0}, \mathbf{\alpha}_1,\dots,\mathbf{\alpha}_K, \mathbf{\beta}_1,\dots,\mathbf{\beta}_K\}$ denotes the set of decision variables, $M$ is a large constant for integral constraints, and $\packageset{}$ in~\eqref{eq:mvnnmilp_packagecons} is the set of feasible packages for the prosumer. A detailed analysis is given in~\cite{Weissteiner_Heiss_Siems_Seuken_2022}.

\begin{algorithm}[t]
    \caption{Learn value function}\label{alg:io}
    \begin{algorithmic}[1]
        \State Input: $\dataset{t}{}:=\{(\dpackage{i}{k}, \linprice{}{k})\}_{k=0}^t$, $\{\alpha_p\}_{p=0}^P$, $P$, $\params{}{0}$
        \Procedure{Train network}{}
            \For{$p\gets \{0,\dots,P\}$}
                \For{$k\gets \{0,\dots,t\}$}
                    \State $\lpackage{i}{k} \leftarrow$ Solve~\eqref{eq:mvnnmilp} with parameters $\linprice{}{k}$ and $\params{}{p}$
                \EndFor
                \State Loss: $l_p \leftarrow \sum_k \lval{i}(\lpackage{i}{k}; \params{}{}) - \lval{i}(\dpackage{i}{k}; \params{}{})$
                \State Gradient step: $\params{}{p+1}\leftarrow \params{}{p} - \alpha_p\nabla_{\params{}{}} l_p$
            \EndFor
        \EndProcedure
        \State Return $\params{}{P}$
    \end{algorithmic}
\end{algorithm}
}
\end{document}